# Respiration monitoring by combining EMG and bioimpedance measurements

Roman Kusche[1][0000-0003-2925-7638], and Martin Ryschka[1]

[1] Laboratory of Medical Electronics, Luebeck University of Applied Sciences, 23562 Luebeck, Germany
`roman.kusche@fh-luebeck.de; martin.ryschka@fh-luebeck.de`

**Abstract.** A common technique to measure diaphragm electrical activation is the acquisition of the occurring electromyography signals using surface electrodes. A significant problem of this technique is its sensitivity against motion artifacts. Forces or vibrations can influence the electrode skin contacts, which generate changes of the electrodes' half-cell voltages as well as the electrode skin impedance. Both effects result in noise in the same frequency range as the typical electromyography signal and therefore it's hard to separate these distortions from the desired signal.

Another technique to detect muscle contractions is the electrical impedance myography. By applying a small alternating current to the tissue of interest and measuring the occurring voltage drop, the bioimpedance is determined. It contains the information about muscle contractions and relaxations. The major advantage of this technique is, that the impedance information is coded as the amplitude modulation of that voltage drop. The frequency of the excitation current is typically chosen in the range of tens of kHz and thus can easily be separated from the above mentioned artefacts.

This work describes a measurement setup which is capable of acquiring electromyography as well as bioimpedance signals simultaneously, sharing the same electrodes. An analog circuit is presented which combines the information of both measurement techniques, allowing their common analog-to-digital conversion by a single converter. The system is capable to acquire 1000 bioimpedances/electromyography samples per second with a resolution of 24 bits.

First measurements show that signal distortions, caused by vibrations at the electrodes, are attenuated significantly in bioimpedance measurements.

**Keywords:** Impedance Myography, Respiration Monitoring, Electromyography, Diaphragm Electrical Activation (EAdi), EMGdi, Bioimpedance.

## 1    Introduction

Each respiration cycle is controlled by the contraction of the diaphragm [1]. Therefore, monitoring of this muscle is of interest in many biomedical applications, which are focused on respiration. A well-known method is acquiring the electrical activation of the diaphragm (EAdi) [2]. This specific electromyography (EMGdi) application can be performed by attaching needle or surface electrodes on the thorax, whereas this work focusses on the usage of surface electrodes.

There are two major challenges regarding EMGdi measurements. First, the heart of the subject is close to the diaphragm. Therefore, there is a superposition of the actual EMGdi voltage signal and the ECG signal. Unfortunately, the high frequency components of the ECG are in the same frequency range as the EMGdi signal [1].

The second problem are the electrode-skin interfaces. The occurring half-cell voltages, which depend on the chosen kind of electrode, vary when mechanical forces occur [3]. Especially vibrations in the frequency range of the EMGdi signal can consequently lead to misinterpretations of the acquired signals.

The approach of this work is to modulate the information about the muscle contraction to a much higher frequency range, by performing bioimpedance measurements. Since the origin of the bioimpedance signal differs from the EMGdi's origin, both signals do not exactly contain the same kind of information [4]. Furthermore, present signal processing algorithms are commonly focused on EMG signals. Due to these facts, it is desirable to measure the EMGdi and the bioimpedance simultaneously.

In this work we propose a measurement approach to monitor respiration via the simultaneous acquisition of the diaphragm's EMG as well as its bioimpedance. Afterwards, the technical implementation and a first measurement from a human subject are presented.

## 2 Materials and Methods

### 2.1 Electromyography

Electromyography is the acquisition of the action potentials of muscles. Therefore, surface electrodes can be used to acquire the occurring differential voltages. Commonly, this signal is in millivolt ranges and has its major frequency components between 10 Hz and 500 Hz [5].

### 2.2 Impedance Myography

Impedance myography is the measurement of changes in the electrical bioimpedance of muscles, caused by geometrical changes. Therefore, a small AC current in a frequency range of several kHz is applied to the tissue under observation via electrodes. The occurring voltage drop over the muscle is amplitude modulated (AM) by the impedance changes within the tissue [4].

### 2.3 Measurement Approach

Both signals, the EMG as well as the bioimpedance signal are electrical signals, which can be acquired by using surface electrodes. Therefore, this approach is based on the usage of a common set of electrodes.

As described above, the frequency ranges of the EMG signal and of the bioimpedance signal are very different. The interesting frequency range of the EMG signal is marked in Fig. 1 (red). The lower end of this range ($f_{EMG,l}$) is typically about 10 Hz, whereas the higher end ($f_{EMG,h}$) can reach up to hundreds of Hz [5]. The carrier frequency of the bioimpedance measurement ($f_c$) is about three decades higher, respectively in ranges of tens or hundreds of kHz. However, the modulating frequency ($f_m$), corresponding to the respiration related impedance changes, is typically below 1 Hz. Therefore, the depicted frequency band of the bioimpedance measurement (blue) is very small in the real application.

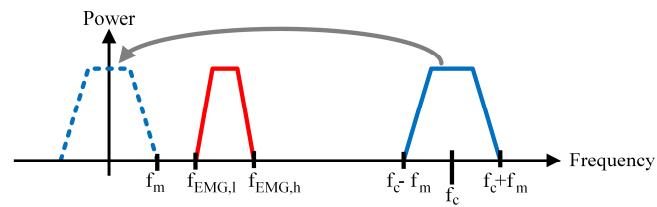

**Fig. 1.** Signal spectrum of the superposition of the EMG signal and the corresponding bioimpedance measurement signal.

To avoid a very high analog-to-digital (ADC) sampling rate and the entailing high computational effort for the AM demodulation, the acquired voltage signal is preprocessed. Since the frequency range of the signal below $f_{EMG,l}$ is unused so far, the narrowband signal of the bioimpedance can be shifted, as shown in Fig. 1. Obviously, this procedure is only applicable when the condition $f_m < f_{EMG,l}$ is fulfilled.

The resulting required sampling frequency can consequently be reduced to $f_s > 2 \cdot f_{EMG,h}$.

### 2.4 Implementation

The proposed approach has been implemented as a mixed signal system. The corresponding block diagram is shown in Fig. 2.

On the left side of the illustration, the tissue under observation is modelled as a series circuit consisting of the bioimpedance $Z_{Bio}$ and an EMG generating voltage source $V_{EMG}$. To measure the bioimpedance, a current source applies a current of $800 \,\mu A$ with a frequency of 143 kHz to the area of interest. These current values comply with the limitations, given by the standard for medical electrical equipment (IEC 60601-1). The superposition of the occurring voltage over the bioimpedance and the EMG voltage is amplified by an analog differential amplifier. Afterwards, the high frequency components of the bioimpedance measurement are shifted down, as described before. Therefore, an AM demodulation circuit is realized by combining an analog rectifier and a low pass filter ($LP_1$). To avoid the disappearance of the low EMG frequencies, a half-wave rectifier is used.

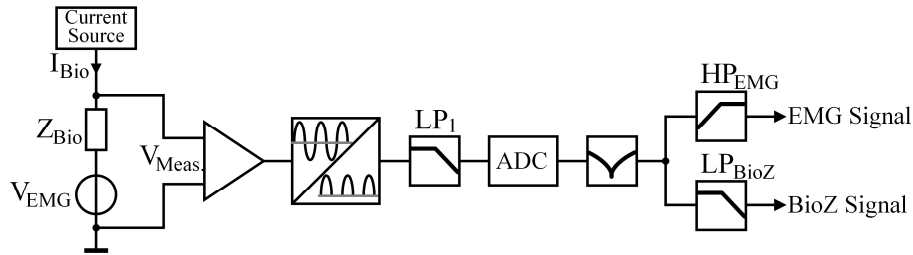

**Fig. 2.** Block diagram of the technical implementation.

The low pass filtered signal is digitized with a sampling rate of 1 kSPS and a resolution of 24 bits. Afterwards, a digital notch filter is used to remove the 50 Hz interference, caused by the mains. In a final step, the EMG signal and the bioimpedance signal are separated from each other. Therefore, a digital high pass filter is used for extracting the EMG signal and a low pass filter to obtain the bioimpedance magnitude information.

## 3 Results

The described measurement instrumentation has been used to perform first measurements on a human subject. In the first experiment, the relationship between the EMG signal and the bioimpedance magnitude is analyzed. Therefore, the measurement setup, shown in Fig. 3, was used. Using a common set of surface electrodes, positioned at the thorax, the bioimpedance $Z_{di}$ and the electromyography signal $EMG_{di}$ have been acquired.

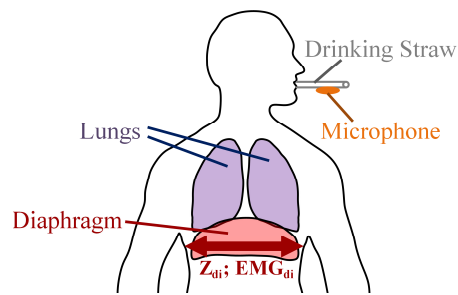

**Fig. 3.** Setup to acquire the respiration via EMG and bioimpedance measurements of the diaphragm. As a time reference, a microphone, mounted on a drinking straw, detects the breathing sounds.

To provide a time reference, the respiration was also detected by an additional microphone, positioned on a drinking straw. The subject was asked to inhale through this drinking straw for three times.

In Fig. 4 (a) the resulting signals are depicted. The first plot shows the microphone signal from the drinking straw and therefore indicates the time periods of inhaling. In the second plot, the acquired signal, consisting of the EMG and the bioimpedance component is shown. Since the amplitude of the EMG component is much lower than that of the bioimpedance, it is hard to recognize in this plot. Below this plot, the separated signal components are displayed.

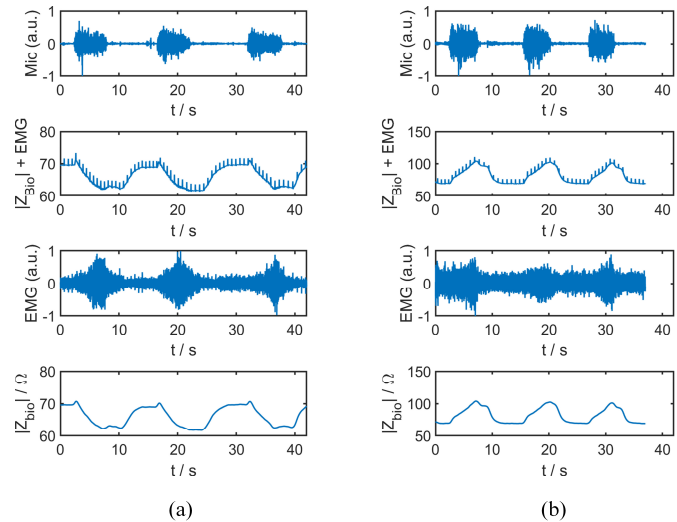

(a)        (b)

**Fig. 4.** Measured signals, acquired via surface electrodes attached on the thorax, close to the diaphragm (a). The signals in (b) have been acquired after relocating the electrodes about 10 cm more upwards.

It can be seen, that during inhaling, when the diaphragm muscle is contracting, the EMG signal increases, whereas the bioimpedance magnitude decreases. This impedance behavior indicates, that actually a muscle contraction is measured [4]. To compare this result with a measurement, more focused on the lungs' impedance changes, the procedure has been repeated with electrodes, attached about 10 cm above the previous position. The results are shown in Fig. 4 (b). It is distinguishable, that the EMG signal is reduced and therefore closer to the noise level. Additionally, now the bioimpedance magnitude increases during inhaling, which corresponds to the typical impedance behavior of the lungs [6].

To compare the robustness against electrode motion artifacts, the experiment has been repeated. The subject was asked to inhale and exhale again. During this cycle, one of the voltage electrodes has been tapped gently. In Fig. 5 the resulting processed signals of the EMG and the bioimpedance are shown for a time span of 7 s. Since the occurring frequency of this kind of interference is much higher than the frequency range of the bioimpedance changes, it can easily be filtered out. However, the EMG signal is disturbed significantly. Some occurring peaks, e.g. at t=3 s or t=5 s, are much higher than the original EMG signal, which can hardly be recognized in this plot. Since the frequency range of the actual EMG signal contains the frequencies of these motion artifacts, common filter techniques are not effective.

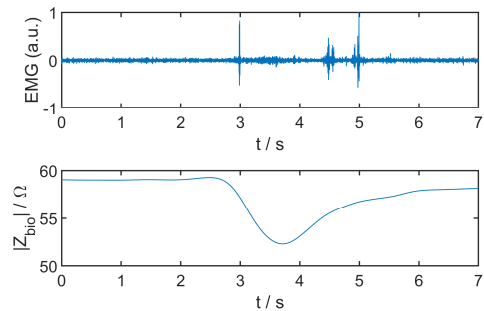

**Fig. 5.** Measurement result of one respiration cycle. To simulate motion artifacts, the positive voltage electrode has been tapped gently at t=3 s, t=4.5 s and t=5 s.

## 4    Discussion

The proposed measurement method has proven to be useful for gathering additional respiration information.

A first experiment has demonstrated the significant influence of the exact electrodes positioning on the measurement results. Therefore, the electrodes positioning has to be performed precisely. Even in the proposed measurement results it is difficult to evaluate the influence of the lungs or other tissue, which changes its geometry during respiration.

The major advantage of performing a bioimpedance measurement is the robustness against disturbances. By using a known modulation frequency, much higher than the expected motion artifacts, and demodulating it to a frequency range below these artifacts, this technique is capable of bypassing the critical frequency band. This beneficial effect has been demonstrated in the second experiment. However, even when distortions in much higher frequency ranges than the respiration cycles are attenuated reliably, there could be additionally distortions in lower frequency ranges, e.g. caused by motion artifacts or drifting effects. These conditions have to be examined in the future.

## Conflicts of Interest

The authors declare that they have no conflict of interest.